\documentclass[twocolumn,preprintnumbers,amsmath,amssymb]{revtex4}

\usepackage{graphicx}
\usepackage{dcolumn}
\usepackage{bm}
\usepackage{amsmath}

\def \be {\begin{equation}}
\def \ee {\end{equation}}

\begin{document}

\title{1/2, 1, and 3/2--Law Non-Radiative Accretion Flows}

\author{Andrei Gruzinov}

\affiliation{ CCPP, Physics Department, New York University, 4 Washington Place, New York, NY 10003
}

\begin{abstract}

Assuming self-similarity of the first kind, we get three possible values $p=1/2,~1,~3/2$ for the exponent describing the density profile, $\rho \propto r^{-p}$, of a non-radiative (and hence quasi-spherical) accretion flow. The high and low $p$ cases are known as Bondi and Convection-Dominated accretion flows. The 1-law flow we tentatively identify with the so-called Magnetically-Frustrated accretion flow. If our interpretation is correct, the accretion flow must be, roughly speaking, a collection of Prandtl's turbulent jets. The 1-law flow, being a first-kind self-similar solution, may actually occur in nature (in collisionless plasma).

\end{abstract}

\maketitle

\section{Introduction}

One of the most fundamental questions of high-energy astrophysics -- how matter falls into black holes -- remains well out of reach of present-day theoretical and numerical methods. For non-radiative (and hence quasispherical) accretion flows this is already clear from the fact that the plasma is collisionless. In what sense is this collisionless plasma describable by magnetohydrodynamics is not known; and purely hydrodynamic description is also manifestly inapplicable as we discuss bellow.

On the other hand, non-radiatively accreting black holes are well-observed (see \cite{Pang2010} for discussion of observations and for references which are needed here and elsewhere); radiation can be dynamically unimportant and yet sufficient to make a black hole visible. One then wants to form a rough idea of at least how the density scales with radius, i.e., what is the exponent $p$ in the formula $\rho \propto r^{-p}$. Once $p$ is known, we get the plasma density at a distance $r$ from the black hole
\be
\rho \sim \rho_0\left({R_B\over r}\right)^p,
\ee
where $\rho_0$ is the density at infinity, $R_B\sim {GM\over c_s^2}$ is the Bondi radius, $M$ is the black hole mass, and $c_s$ is the sound speed at infinity.

The exponent $p$ also gives the rate of accretion. Since near the Schwarzschild radius the gas must flow in at about the speed of light, the accretion rate is 
\be
\dot{M}=\dot{M}_B\left({R_S\over R_B}\right)^{3/2-p},
\ee
where $R_S\sim {GM\over c^2}$ is the Schwarzschild radius and
\be
\dot{M}_B\sim c_s\rho_0R_B^2
\ee
is the canonical Bondi accretion rate.

Knowing the density near the black hole one can try to estimate the observables: luminosity, spectrum, polarization, etc... Thus calculating the exponent $p$ is highly desirable for any discussion of the physics of black hole radiation.

At first sight the Bondi's value $p=3/2$ is the only reasonable choice. Indeed, at about the Bondi radius the free-fall velocity becomes comparable to the thermal velocity, or, to put it differently, gravity overwhelms the pressure support. The plasma has no choice but to flow in at about the free-fall velocity, which is $\sim c_s$ at $r\sim R_B$. One then gets the Bondi accretion rate, $\dot{M}\sim \dot{M}_B$, and hence $p=3/2$. 

Observations seem to disagree with this conclusion. Also, a very powerful argument of Victor Shvartsman \cite{Shvartsman1971} seems inescapable. Shvartsman's argument can be formulated as follows. Assume Bondi flow, but put in some random magnetic field $B$. Assume that $B$ is energetically subdominant at Bondi radius. The laminar Bondi flow is characterized by a high degree of ``spaghettization''. As plasma flows in, the tangential distances between fluid particles decrease as $a_\perp \propto r$, and then, to keep $\rho \propto r^{-3/2}$, the radial distances must increase as $a_r \propto r^{-1/2}$. This ``spaghettization'' will strongly amplify radial $B$. By flux conservation, $Ba_\perp^2=const$, giving $B\propto r^{-2}$, and magnetic energy density $B^2\propto r^{-4}$, which will soon become much larger then the Bondi flow's energy density $\propto r^{-1}\rho\propto r^{-5/2}$. To continue the fall, the plasma must get rid of the magnetic energy. This leads to an entropy increase with decreasing radius; assuming that the temperature always scales as $T\propto r^{-1}$, the requirement of increasing entropy gives $p<3/2$.

We offer an estimate of the actual value of $p$ in \S\ref{self}.

\section{First-kind self-similar accretion flow}\label{self}

As the gravitational potential, $\propto r^{-1}$, is self-similar, and the plasma equations of motion (be it hydro, magnetohydro, or Vlasov at relevant length scales) are self-similar too, it makes sense to assume that the accretion flow is self-similar. Then all velocities (flow velocity, free-fall velocity, thermal speed, Alfven speed) scale as $r^{-1/2}$. The mass flux $\Phi$, the momentum flux $F$ \footnote{Momentum is, of course, not conserved, it is changed by the gravitational force. However, since all our velocities are of the order of free-fall, momentum may (or may not) be ``scaling-wise'' conserved. For instance if one is ejected at twice the free-fall velocity, the momentum will remain approximately (scaling-wise) constant with changing $r$.}, and the energy flux $L$ scale as
\be
\Phi \propto r^2\rho v \propto r^{3/2-p},
\ee
\be
F \propto r^2(\rho v)v \propto r^{1-p},
\ee
\be
L \propto r^2(\rho v^2)v \propto r^{1/2-p}.
\ee

Assuming self-similar solution of the first kind \cite{Barenblatt1996}, we get the following three possibilities:
\begin{itemize}

\item ${\bf p=3/2,~~\Phi=const,~F=0,~L=0.}$ This is the Bondi flow. Plasma everywhere falls down at about $v$, and the momentum and energy fluxes are kept constant due to precise cancellation of positive and negative contributions. As we have argued, Bondi flow is unphysical (in the non-radiative case).

\item ${\bf p=1/2,~~\Phi=0,~F=0,~L=const.}$ This flow can be thought of as a strongly (sonically) convective nearly-non-accreting atmosphere. It is characterized by strong entropy inversion, which explains vigorous convection. It was thought that, given favorable initial conditions, this flow might be realizable as Convection-Dominated accretion. However, direct numerical simulation does not find it (see \cite{Pang2010} for discussion and references).

\item ${\bf p=1,~~\Phi=0,~F=const,~L=0.}$ This flow might be thought of as a collection of Prandtl's turbulent jets which are known to keep the momentum flux constant, while increasing the mass flow $\Phi$ by entrainment, and decreasing the power $L$ by stirring the surrounding plasma \cite{Landau1987}. As each of the jets propagates out, it gives energy to the incoming plasma, thereby increasing the entropy of the incoming gas. The jet also entrains the incoming gas, thereby decreasing the inflow rate. The flow of \cite{Pang2010} has $p$ close to 1. 

\end{itemize}

\section{Conclusions}

It appears that the numerical simulation \cite{Pang2010} can be explained as the 1-law accretion flow -- a collection of turbulent jets. If so, the value of the exponent found by \cite{Pang2010} may remain valid even for collisionless plasma, as $p=1$ is one of just a discrete set of possible exponents allowed by self-similarity of the first kind.
 
If, on the other hand, better numerics convincingly demonstrates that $p$ is not equal to one, meaning that Magnetically-Frustrated flow is a second-kind self-similar solution, then going to collisionless plasma is likely to change the value of $p$.


\begin{thebibliography}{99}

\bibitem{Pang2010}
Bijia Pang, Ue-Li Pen, Christopher D. Matzner, Stephen R. Green, Matthias Liebendörfer, arXiv:1011.5498 (2010)

\bibitem{Shvartsman1971}
Shvartsman, V. F., Soviet Astronomy - AJ {\bf 15}, 377 (1971)

\bibitem{Barenblatt1996}
G. I. Barenblatt, Scaling, Self-similarity, and Intermediate Asymptotics, Cambridge University Press (1996)

\bibitem{Landau1987}
L. D. Landau, E. M. Lifshitz, Fluid Mechanics, Pergamon Press (1987)


\end{thebibliography}
\end{document}